\documentclass[12pt]{iopart}

\usepackage{iopams}  
\usepackage{graphicx}

\begin{document}

\title{Towards Quantum Supremacy with Lossy Scattershot Boson Sampling}

\author{Ludovico Latmiral$^{1}$, Nicol\`o Spagnolo$^{2}$, Fabio Sciarrino$^{2}$}
\address{$^{1}$QOLS, Blackett Laboratory, Imperial College London, London SW7 2BW, United Kingdom}
\address{$^{2}$Dipartimento di Fisica - Sapienza Universit\`{a} di Roma, P.le Aldo Moro 5, I-00185 Roma, Italy}
\vspace{10pt}

\begin{abstract}
Boson Sampling represents a promising approach to obtain an evidence of the supremacy of quantum systems as a resource for the solution of computational problems. The classical hardness of Boson Sampling has been related to the so called \textit{Permanent-of-Gaussians Conjecture} and has been extended to some generalizations such as Scattershot Boson Sampling, approximate and lossy sampling under some reasonable constraints. However, it is still unclear how demanding these techniques are for a quantum experimental sampler. Starting from a state of the art analysis and taking account of the foreseeable practical limitations, we evaluate and discuss the bound for \textit{quantum supremacy} for different recently proposed approaches, accordingly to today's best known classical simulators.
\end{abstract}

%\maketitle

\section{Introduction} 
The \emph{Boson Sampling} (BS) problem is a well-built example of a dedicated issue that cannot be efficiently solved through classical resources (unless the collapse of polynomial hierarchy to its third level), though it can be tackled with a quantum approach \cite{Aaronson10}. More specifically, it consists in sampling from the probability output distribution of \emph{n} non-interacting bosons evolving through a $m\times m$ unitary transformation. Together with applications in quantum simulation \cite{huh2015} and searching problems \cite{Aaronson10b}, the aim of a Boson Sampling device is to outperform its classical simulator counterpart. This would provide strong evidence against Extended Church-Turing Thesis and would represent a demonstration of quantum supremacy\footnote{Extended Church-Turing Thesis conjectures that \emph{a probabilistic Turing machine can efficiently simulate any realistic model of computation}, where \emph{efficiently} means up to polynomial-time reductions.}.
Following the initial proposal, many experiments have been settled so far by using linear optical interferometers \cite{Broome2013, Spring2013, crespi2012, Tillmann2012} where indistinguishable photons are sent in an interferometric lattice made-up of passive optical elements such as beam splitters and phase shifters. In the perspective of implementing a scalable device, one of the main differences with respect to a universal quantum computer is that only passive operations are permitted before detection. This implies that it is not known whether it is possible to apply quantum error correction and fault tolerance \cite{Aharonov1997, Knill1998,Rohde2014}.

This apparent limitation was already considered in the first proposal \cite{Aaronson10}, where the problem was proved to be classically hard also lowering the demand to approximate Boson Sampling, under mild constraints. Many papers have focused on this issue \cite{Leverrier2015} as well on several possible causes of experimental errors \cite{Rohde2012, Rohde2012a, Rohde2014, Shchesnovich14_2, Shchesnovich2015, Tillmann14}. The intensive discussion on this topic has triggered a number both of theoretical \cite{Gogolin2013, Aaronson13,Tichy2013,Aolita15,Walschaers2014,Shch2016,bentivegna2016} and experimental \cite{Spagnolo2014, Carolan2014, crespi2015, carolan2015} studies on the validation of a Boson Sampler, i.e. the assessment that the output data sets are not generated by other efficiently computable models.  Moreover, an advantageous variant of the problem called \textit{Scattershot} Boson Sampling has been theoretically proposed \cite{Lund2013, Aaronsonblog} and experimentally implemented \cite{bentivegna2015} in order to better exploit the peculiarities of the experimental apparatus based on spontaneous parametric down conversion (SPDC). It was eventually very recently proved that the same hardness result holds when there is a constant number of photons lost before being input, which in turn can presumably be extended to constant losses at the output \cite{Aaronson2015}.

In this paper we review the fundamental issue of experimental limitations to understand which are the requirements that make the implementation suitable to reach quantum supremacy. We define the latter as the regime where the quantum agent samples faster than his classical counterpart. We analyze the state of the art together with all the complexity requirements, reviewing the whole process in light of recent theoretical extensions and experimental proposals \cite{Aaronson2015, huh2015bis}. Starting from the already established idea of sampling with constant losses occurring only at the input, we discuss the extension of Boson Sampling to a more general lossy case, where photons might be lost either at the input and/or at the output. This method provides a gain from the experimental perspective both in terms of efficiency and of effectiveness. Indeed, we actually estimate a new threshold for the achievement of quantum supremacy and we show how the application of such generalizations could pave the way towards beating this updated bound.

\section{Standard and Scattershot Boson Sampling}
\label{sec:BS}
Boson Sampling (BS) consists in sampling from the probability distribution over the possible Fock states $\vert T \rangle$ of $n$ indistinguishable photons distributed over $m$ spatial modes, after their evolution through a $m \times m$ interferometer which operates a unitary transformation $U$ on their initial, known, Fock state $\vert S \rangle$. If $s_i$ ($t_j$) denotes the occupation number for mode \emph{i} (\emph{j}), the transition amplitude from the input to the output configuration is proportional to the permanent of the $n \times n$ matrix $U_{S,T}$ obtained by repeating and crossing $s_i$ times the $i^{\mathrm{th}}$ column with $t_j$ times the $j^{\mathrm{th}}$ row of $U$ \cite{Scheel04}
\begin{equation}
\label{bs_equation}
\langle T \vert U_{\mathrm{F}}\vert S \rangle = \frac{\mathrm{per}(U_{S,T})}{\sqrtsign{s_{1}!\dots s_{m}!t_{1}!\dots t_{m}!}},
\end{equation}
where $U_{\mathrm{F}}$ represents the associated transformation on the Fock space. Given a square matrix $A_{n\times n}$, its permanent is defined as $\mathrm{per}(A)=\sum_{\sigma}\prod_{i=1}^n a_{i,\sigma(i)}$, where the sum extends over all permutations of the columns of $A$. If $A$ is a complex (Haar) unitary the permanent is {\bf \#P}-hard even to approximate \cite{Valiant79}. Conversely, for a nonnegative matrix it can be classically approximated in probabilistic polynomial time \cite{Jerrum2004}. The most efficient way to compute the permanent of a $n\times n$ matrix $A$ with elements $a_{i,j}$ is currently Glynn's formula \cite{glynn2010}
\begin{equation}
\mathrm{per}(A)=\left[\sum_\delta (\prod_{k=1}^n\delta_k)\prod_{j=1}^n\sum_{i=1}^n\delta_ia_{i,j}\right]\cdot 2^{1-n},
\end{equation}
where the outer sum is over all possible $2^{n-1}$ $n$-dimensional vectors $\vec{\delta}=(\delta_1=1,\delta_2,\cdots\delta_n)$ with $\delta_{i\neq 1}\in\{\pm 1\}$. Processing these vectors in Gray code (i.e. changing the content of only one bit per time, so that the number of counting operations is minimized to $O(n)$) allows the number of steps to scale as $O(n\, 2^n)$.

\begin{figure}[ht!]
\centering
\includegraphics[width=0.99\textwidth]{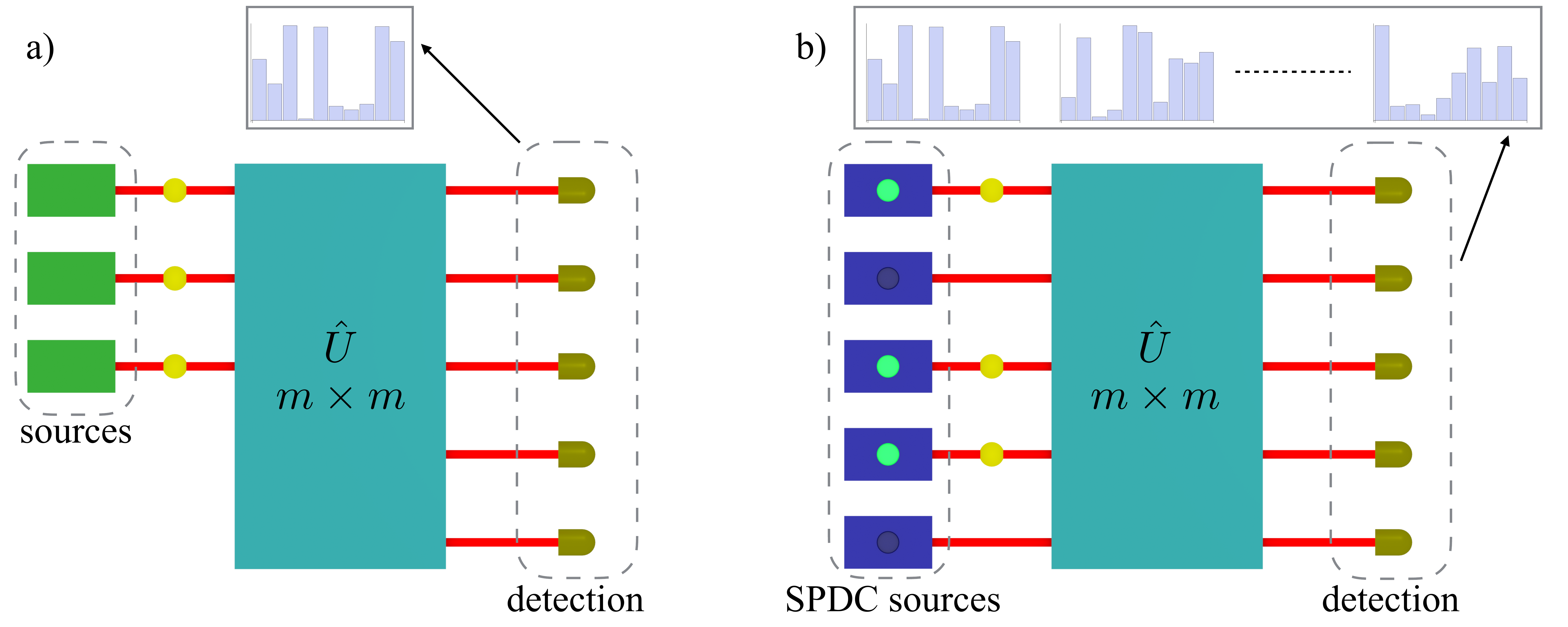}
\caption{a) Conventional Boson Sampling: the linear transformation is sampled with $n$ sources, injecting a fixed input state for each run. b) Scattershot Boson Sampling: $m$ SPDC sources are connected in parallel to the $m$ ports of a linear transformation. Each event is sampled from a random (though known) input state.}
\label{scattershot_conc}
\end{figure}

While in the original proposal all the samples are derived from the same input, \textit{Scattershot} Boson Sampling consists in injecting each time a random, though known, input state. To this end, each input mode of a linear interferometer is fed with one output of a SPDC source (see Fig. \ref{scattershot_conc}). Successful detection of the corresponding twin photon heralds the injection of a photon in a specific mode of the device. It has been proved that the Scattershot version of the BS problem still maintains at least the same computational complexity of the original problem \cite{Lund2013, Aaronsonblog}.
Since BS was proved to be hard only in the regime $m \gg n^{2}$, attention can be restricted only to those $m \choose n$ outputs with no more than one photon per mode among all possible $m+n-1 \choose n$ output states. This also helps to overcome the experimental difficulty to resolve the number of photons in each output mode.

\begin{figure}[ht!]
\centering
\includegraphics[width=0.6\textwidth]{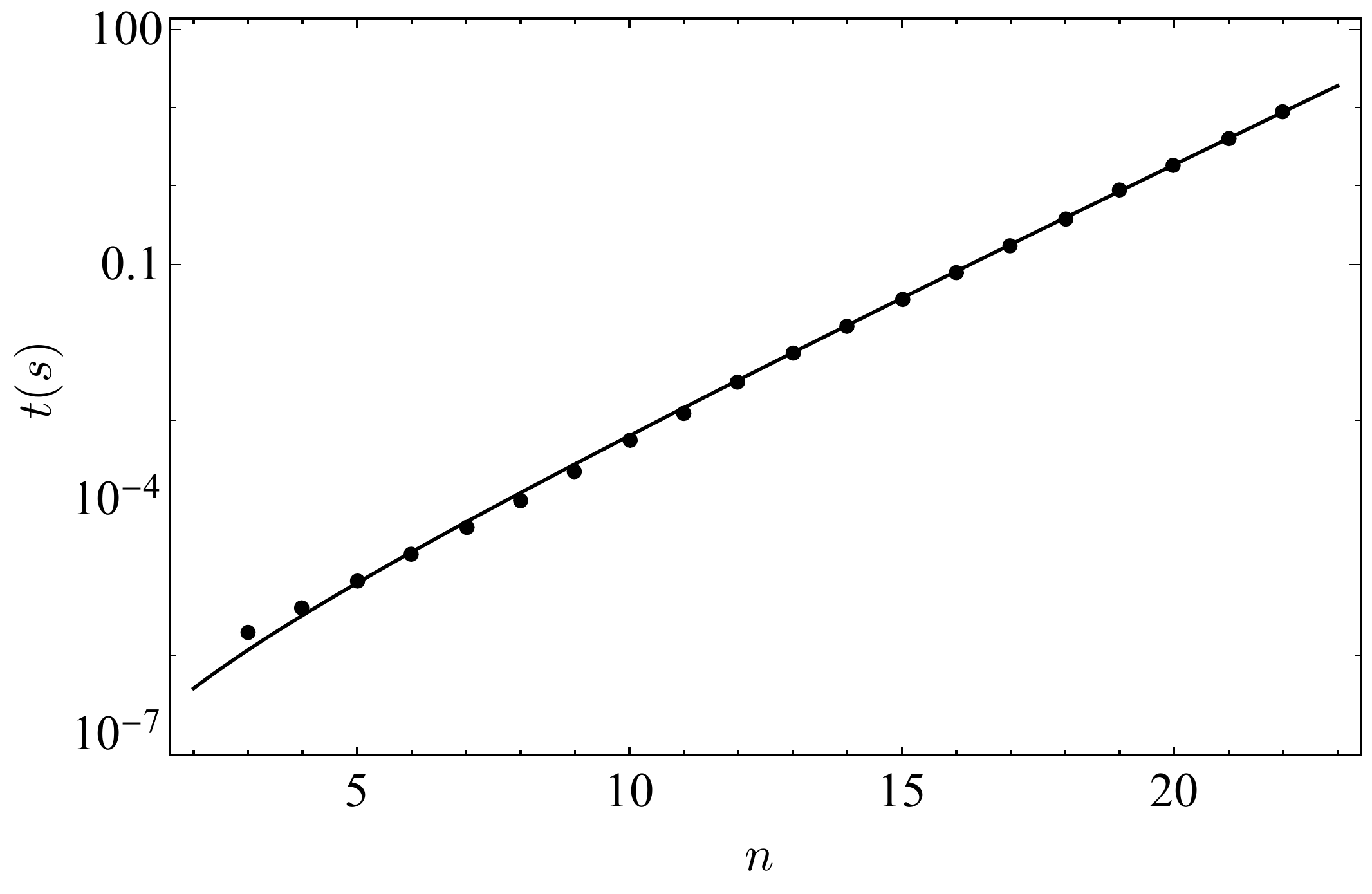}
\caption{Computer simulations of the time required to compute permanents of different size $n$ on a 4 cores 2.3 GHz processor. The fitting function is of the form $A\, n\, 2^{B n}$, with $A=4.47\times 10^{-8}$ and $B=1.05$: the fact that $B$ is slightly greater than one can be explained by the exponential increase in terms of memory resources. The time required for the complete calculation of a boson sampling output probability distribution of $n$ photons in $m$ modes will scale as ${m \choose n} A\, n\, 2^{B n}$.}
\label{permanents}
\end{figure}

To give an idea of the computational complexity behind the BS problem, we show in Fig.\ref{permanents} the real time an ordinary PC requires to calculate a permanent of various size. The time needed to perform exact classical calculation of a complete BS distribution is enhanced by a factor $m \choose n$. The values for the most powerful existing computer, which is approximately one million times faster, can be obtained by straightforward calculations. 
Currently no other approaches different from a brute force simulation, that is, calculation of the full distribution and (efficient) sampling of a finite number of events, have been reported in the literature to perform the classical simulation of BS experiments with a general interferometer.

\section{Scattershot Boson Sampling in lossy conditions}
\label{sec:SBS}

We are now going to discuss how a Scattershot BS experiment with optical photons depends on the parameters of the setup. We will analyze how errors in the input state preparation and system's inefficiencies (i.e. losses and failed detections) affect the scalability of the experimental apparatus. We will not consider here issues such as photons partial distinguishability and imperfections in the implementation of the optical network, since in certain conditions they do not affect the scalability of the system. For the input state, the average mutual fidelity of single photons must satisfy $1-\langle F\rangle\sim O(n^{-1})$ \cite{Shchesnovich2015,Tillmann14}. Necessary conditions in terms of fidelity $F_{\mathrm{el}}=1-O(n^{-2})$ \cite{Leverrier2015} and sufficient conditions in terms of operator distance $\vert \vert A - \tilde{A} \vert \vert_{\mathrm{op}} = O(n^{-2}/\log m)$ \cite{Arkh2015} have been also investigated for the amount of tolerable noise on the network optical elements.

\begin{figure}[ht!]
\centering
\includegraphics[width=0.99\textwidth]{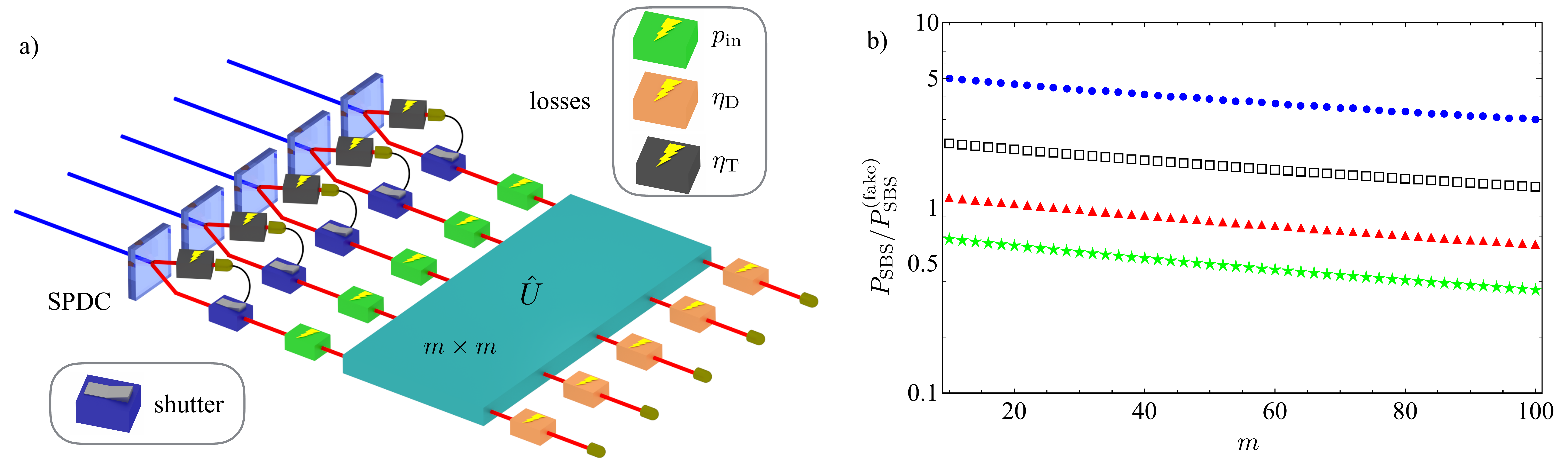}
\caption{a) Schematic view of Scattershot BS, consisting in connecting many parallel SPDC sources to different input modes of the interferometer and post-selecting on the heralded photons. Optical shutters are placed before the input modes to avoid photon injection into wrong ports (i.e. without proper heralding). Losses are divided in $\eta_{\mathrm{T}}$ (single-photon triggering probability), $p_{\mathrm{in}}$ (injection losses) and $\eta_{\mathrm{D}}$ (detection losses). b) Probabilities to successfully carry on a correct Scattershot BS experiment,  i.e. to sample from the single photon Fock states corresponding to those heralded by the triggers, expressed by the ratio $P_{\mathrm{SBS}}/P^{(\mathrm{fake})}_{\mathrm{SBS}}$. The probability decreases if we increase the number of modes and photons: blue circles $n=4$, black squares $n=6$, red triangles $n=8$ and green stars $n=10$. Experimental parameters are set as: $g=0.02$, $\eta_{\mathrm{T}}=0.6$, $p_{\mathrm{in}}=0.7$ and $\eta_{\mathrm{D}}=0.6-0.25*(m-10)/90$ (the probability for a photon to propagate through the interferometer and to be finally detected decreases when we increase the dimension).}
\label{scattershot}
\end{figure}

Spontaneous parametric down conversion is the most suitable known to-date technique to prepare optical heralded single-photon states. Photon pairs are emitted probabilitistically in two spatial modes, and one of the photons is measured to witness the presence of the twin photon. Note that without post-selecting upon the heralded photons, the input state would be Gaussian and thus the distribution would not be hard if detected with a system performing Gaussian measurements \cite{Bartlett2003, Gogolin2013}. The main drawback of using SPDC sources is in the need of a compromise between the generation rate and the multiple pair emission. Indeed, the single-pair probability $g$ has to be kept low so as to avoid the injection of more than two photons in the same optical mode. Hence, it reveals to be essential to consider at least the noise introduced by second order terms that characterize double pairs generation which scales as $\sim g^2$ (see Appendix A for additional information). The probability for $m$ SPDC sources in parallel to generate $s$ single pairs and $t$ double pairs will hence read
\begin{equation}
P_{\mathrm{gen}}^{(2)}(s,t)=g^sg^{2t}(1-g-g^2)^{m-s-t}{m \choose {s,t}},
\end{equation}
where ${m \choose {s,t}}$ is the multinomial coefficient $m!/((m-s-t)!s!t!)$. This expression includes all possible combinations ${m \choose {s,t}}$ of $s$ sources generating one pair ($g^{s}$) and $t$ sources generating two pairs ($g^{2t}$). We show in Fig. \ref{scattershot}a a schematic representation of a Scattershot BS setup where we depicted all the experimental parameters that we define below.
We define with $\eta_{\mathrm{T}}$ the probability to trigger a single photon, leaving out dark counts. If we assume that we do not employ photon number resolving detectors (accordingly with the performance of current technology), the probability that a detector clicks with $n$ input photons is then given by $1-(1-\eta_{\mathrm{T}})^n$. Meanwhile, we call $p_{\mathrm{in}}$ the probability that a single photon is correctly injected in the interferometer, while $\eta_{\mathrm{D}}$ is the probability that the injected photon does not get lost in the network and is eventually detected at the output.

In addition to the original scheme for Scattershot Boson Sampling, optical shutters, that is, a set of vacuum stoppers, are placed on each of the $m$ input modes. The shutters are open only in presence of a click on the corresponding heralding detector, thus ruling out the possibility of injecting photons from unheralded modes. The hypothesis of working in a post-selected regime (with shutters) is helpful in this context: indeed, we are interested only in those events where exactly $n$ photons enter and exit the chip, disregarding every other possible combination.
After some combinatoric manipulation, we derive the probability to successfully perform a Scattershot BS experiment with $n$ photons (i.e. an experiment where $n$ triggers click, $n$ single photons are injected and successfully detected at the output)
\begin{eqnarray}
\fl P_{\mathrm{SBS}}(n)=\eta_{\mathrm{D}}^n\sum_{q=n}^m\sum_{t=0}^q P_{\mathrm{gen}}^{(2)}(q-t,t)\sum_{n_1=\max[n-t,0]}^{\min[q-t,n]}(p_{\mathrm{in}}\eta_{\mathrm{T}})^{n_1} [2p_{\mathrm{in}}(1-p_{\mathrm{in}})\eta_{\mathrm{T}_{2}}]^{n-n_1} \nonumber \\ \times(1-\eta_{\mathrm{T}})^{q-t-n_1}{q-t \choose n_1} (1-\eta_{\mathrm{T}_{2}})^{t-n+n_1}{t \choose {n-n_1}},
\end{eqnarray}
where $\eta_{\mathrm{T}_{2}}$ is the probability to detect a pair of photons $\eta_{\mathrm{T}_{2}} = [1-(1-\eta_{\mathrm{T}})^2]$. The outer sums consider all possible Scattershot single photon and pairs generations, while the inner sum constraints the number of correctly injected single photons to $n$ (among these, only $n_1$ derive from single generated pairs).

However, from an experimental point of view we only know that $n$ detectors have clicked both at the input and at the output. Hence, we cannot rule out that this was the result of a fake sampling where additional photons have been injected and some erroneous compensation has occurred (e.g. unsuccessful injection of single photons, losses in the interferometer, failures in the output detection). Indeed, the probability to carry out an experiment from an (non-verifiable) incorrect input state is given by
\begin{equation}
P^{(\mathrm{fake})}_{\mathrm{SBS}}(n)=\sum_{q=n}^m\sum_{t=1}^q P_{\mathrm{trig,det}}^{(\mathrm{fake})}(n|(q-t),t)
\end{equation}
where we sum the probability to inject a fake state, while triggering and detecting $n$ photons, over all possible generations with $t$ double pairs: $P_{\mathrm{trig,det}}^{(\mathrm{fake})}(n|(q-t),t)$ (see Appendix B for full details on the calculation).

We plot in Fig. \ref{scattershot}b a numerical analysis of the ratio $P_{\mathrm{SBS}}/P^{(\mathrm{fake})}_{\mathrm{SBS}}$ for different numbers of photons, varying the number of modes and accordingly changing the detection probability $\eta_{\mathrm{D}}$ in a feasible way. We obtain in parallel that the ratio of correctly sampled events over the fake ones is highly dependent on the number of extra undetected photons. Indeed, this ratio is actually a decreasing function of $g$ and $p_{\mathrm{in}}$, since higher values of these parameters increase the weight of multiphoton emission and injection, and an increasing function of $\eta_{\mathrm{T}}$ and $\eta_{\mathrm{D}}$. 

\section{Validation with losses}
\label{sec:valid}
We will discuss here some extensions of the system that could boost quantum experiments towards reaching the classical limit. A major contribution in this direction came from Scott Aaronson and Daniel Brod who generalized BS to the case where a constant number of losses occurs in input \cite{Aaronson2015}, though setting the stage for losses at the output as well. Addressing their proposals, we discuss here the problem of successfully validating these lossy models against the output distribution of distinguishable photons, representing a significant benchmark to be addressed. Indeed, it is still an open question whether it is possible to discriminate true multiphoton events with respect to data sampled from easy-to-compute distributions. A non-trivial example is given by the output distribution obtained when the same unitary is injected with distinguishable photons. The latter presents rather close similarities with the true BS one, and at the same time provides a physically motivated alternative model to be excluded. A possible approach to validate BS data against this alternative hypothesis is a statistical likelihood ratio test \cite{Spagnolo2014,Bentivegna2014Valid}, which requires calculating the output probability assigned to each sampled event by both the distributions (i.e. a permanent). In this case a validation parameter $\mathcal{V}$ is defined as the product over a given number of samples of the ratios between the probabilities assigned to the occurred outcomes by the BS distribution and the distinguishable one. The certification is considered successful if $\mathcal{V}$ is greater than one with a 95\% confidence level after a fixed number of samples. On one side, the number of data required to validate scales inversely with the number of photons and is constant with respect to the modes. This means that with this method there is no exponential overhead in terms of number of necessary events. Conversely, the need of evaluating matrix permanents to apply the test implies an exponential (in $n$) computational overhead.

A relevant question is then if lossy Boson Sampling with indistinguishable photons can in principle be discriminated from lossy sampling with distinguishable particles. The same likelihood ratio technique can be adopted to validate a sample in which some losses have occurred. Indeed, for each event we apply the protocol by including in each output probability all the cases that could have yielded the given outcome. This calculation is performed both in the BS and in the distinguishable photons picture.
We thus verified that the scaling in $n$ and $m$ obtained in the lossless case is preserved when constant losses in input are considered, that is, $n^{\mathrm{in}}_{\mathrm{lost}}$ constant with respect to $n$. We will then show in Sec.\ref{quantumsupremacybound} that constant losses still boost the system performances. 

\begin{figure}[ht!]
\centering
\includegraphics[width=0.99\textwidth]{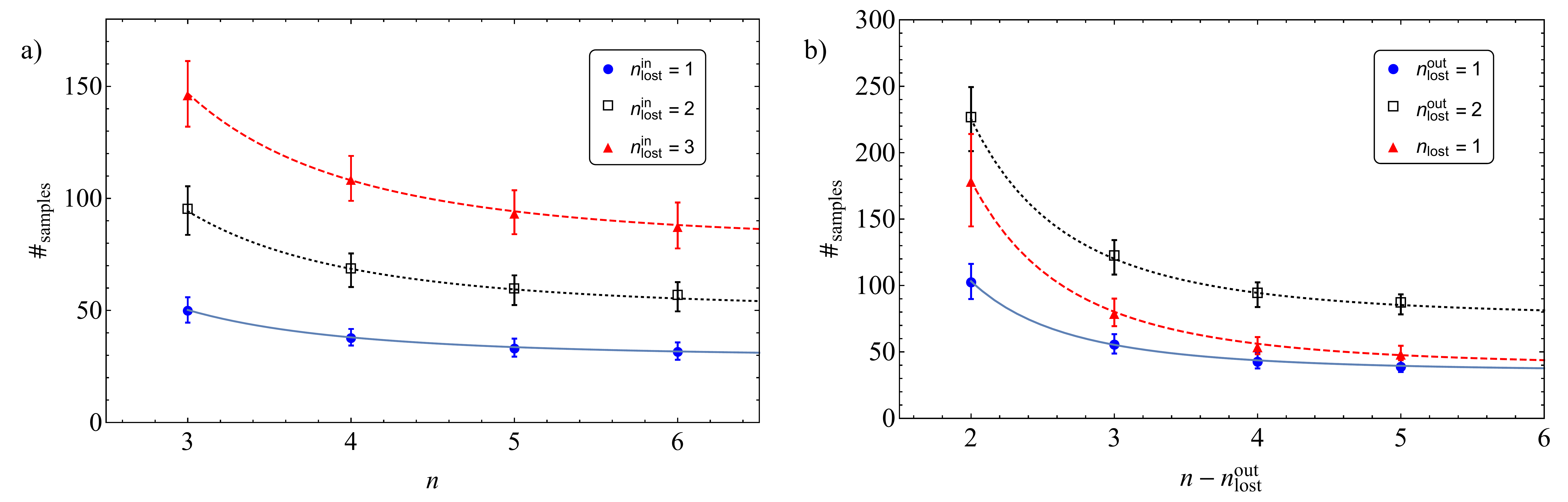}
\caption{Minimum data set size to validate lossy Boson Sampling against a sampling with distinguishable photons with a 95\% confidence level. The results have been averaged over $100$ Haar random $30\times30$ unitaries, though they are almost independent of the dimension within the regime $m>n^2$ (see Appendix C). a) Losses occur only at the input: $n+n^{\mathrm{in}}_{\mathrm{lost}}$ are triggered but only $n$ photons are injected and finally detected at the output. The number of samples decreases as $\#_{\mathrm{samples}}=A+B\, n^{-3}$ for fixed $n^{\mathrm{in}}_{\mathrm{lost}}$ and increases with the losses (vertically aligned data). b) Losses occur at the output: $n$ photons are triggered and injected, but only $n-n_{\mathrm{lost}}^{\mathrm{out}}$ are detected ($n_{\mathrm{lost}}^{\mathrm{out}}=1$ blue circles and $n_{\mathrm{lost}}^{\mathrm{out}}=2$ black squares). The red triangles represent the case in which one photon can be lost with equal probability either at the input or at the output. The number of samples necessary to validate decreases as $\#_{\mathrm{samples}}=A+B\,\tilde{n}^{-3}$, where $\tilde{n}$ is the number of detected photons.}
\label{lossyBS}
\end{figure}

Additionally, we have considered the case where losses happen at the output, after the evolution, and the combined case when they might occur both at the input and at the output. We plot in Fig. \ref{lossyBS}b the validation of a $30$ modes BS device for these lossy cases, verifying the scalability with respect to the number of photons. This result confirms the findings of \cite{Aaronson2015} that constant losses with respect to the number of photons should not affect the complexity of the problem. It is thus a relevant basis for the definition of a new problem, \textit{lossy} Scattershot Boson Sampling, which, as we are going to show, allows to lower the bound for quantum supremacy.

\section{The bound for Quantum Supremacy}
\label{quantumsupremacybound}
We can now discuss a threshold for quantum supremacy by resuming all the considerations and the experimental details related to the implementation of Scattershot BS with optical photons that we have presented so far, including losses at the input and at the output. Let us call $t_{\mathrm{c}}$ the time required to classically sample a single BS event and $t_{\mathrm{q}}$ the one for a successful experimental run, our aim is then to calculate the set of parameters that define the region where $t_{\mathrm{c}}/t_{\mathrm{q}} > 1$. As discussed in Sec. \ref{sec:BS}, if $m$ is the number of modes, the time required by a classical computer to simulate a single Scattershot BS run with $n$ photons by using a brute force approach (classical computation of the full distribution and efficient sampling of an output event) is given by:
\begin{equation}\label{classtime}
t_{\mathrm{c}}(m,n)=A^\prime\, n \, 2^{n} {m \choose n},
\end{equation}
where $A^\prime\sim 1.2 \times 10^{-14}s$ is the estimated time scaling for \textit{Tianhe 2}, the most efficient existing computer capable of $34$ petaFLOPS (a first run with $A^\prime\sim 6 \times 10^{-14}s$ has been recently reported in \cite{wu2016}). On the other hand, a quantum competitor that arranges $m$ single photon sources connected in parallel to $m$ inputs could theoretically sample from any event with $n\leq m$ photons. However, runs with too many or too few photons will be strongly suppressed: in particular, we will have to wait on average
\begin{equation}\label{quanttime}
t_{\mathrm{q}}(m,n)=[F_{\mathrm{pump}}^{\mathrm{rate}} (P_{\mathrm{SBS}}(m,n)+\sum_{n_{\mathrm{lost}}}P_{\mathrm{SBS}}^{\mathrm{lossy}}(m,n,n_{\mathrm{lost}}))]^{-1}
\end{equation}
to sample from a $n$ photons generalized Scattershot BS run, i.e. either a successful or a lossy experiment. Indeed, $F_{\mathrm{pump}}^{\mathrm{rate}}$ is the rate at which the laser pumps photons in the SPDC sources, $P_{\mathrm{SBS}}(m,n)$ is the probability to correctly perform a $n$ photons BS given $m$ sources and $P_{\mathrm{SBS}}^{\mathrm{lossy}}(m,n,n_{\mathrm{lost}})$ reads
\begin{eqnarray}
\label{SBSlossy}
\fl P_{\mathrm{SBS}}^{\mathrm{lossy}}(m,n,n_{\mathrm{lost}})=\sum_{i=0}^{n_{\mathrm{lost}}}\eta_{\mathrm{D}}^{n-n_{\mathrm{lost}}}(1-\eta_{\mathrm{D}})^{n_{\mathrm{lost}}-i}{{n-i} \choose {n_{\mathrm{lost}}-i}}\sum_{q=n}^m\sum_{t=0}^q \Bigg[P_{\mathrm{gen}}^{(2)}(q-t,t) \nonumber\\ 
\times \sum_{j=0}^{i}\sum_{n_1=\max[n-t,0]}^{\min[q-t,n]}2^{n-n_1-i+j} p_{\mathrm{in}}^{n-i}(1-p)^{n+i-n_1}\eta_{\mathrm{T}}^{n_1}\eta_{\mathrm{T}_{2}}^{n-n_1} \nonumber
\\ \times(1-\eta_{\mathrm{T}})^{q+t+n_1-2n} {n_1 \choose j} {{n-n_1}\choose {i-j}}{{q-t} \choose {n_1}}{t \choose {n-n_1}}\Bigg],
\end{eqnarray}
%with $i=n_{\mathrm{lost}}^{\mathrm{in}}$ the photons lost at the input and $j$ the fraction of lost photons coming from correctly generated single pairs. 
In this expression, we consider all possible cases where $q-t$ single pairs and $t$ double pairs are generated, $n$ trigger detectors successfully click ($n_{1}$ single-photon inputs with detection probability $\eta_{\mathrm{T}}$, $n-n_{1}$ two-photon inputs with detection probability $\eta_{\mathrm{T}_{2}}$), $i=n_{\mathrm{lost}}^{\mathrm{in}}$ photons are lost at the input (each one with efficiency $p_{\mathrm{in}}$), $j$ is the fraction of lost photons coming from correctly generated single pairs, and finally $n-n_{\mathrm{lost}}$ photons are detected at the output (each one with detection efficiency $\eta_{\mathrm{D}}$).

As we have just shown, $P_{\mathrm{SBS}}$ and $P_{\mathrm{SBS}}^{\mathrm{lossy}}$ depend on the experimental parameters such as the detectors efficiency, the coupling among various segments in the interferometer and the single photon sources. If $n_{\mathrm{lost}}$ is the difference between the number of heralded and detected photons, the probability of a lossy BS with $n-n_{\mathrm{lost}}$ photons will be the sum of all possible cases in which $n_{\mathrm{lost}}=n_{\mathrm{lost}}^{\mathrm{in}}+n_{\mathrm{lost}}^{\mathrm{out}}$, where $n_{\mathrm{lost}}^{\mathrm{in}}$ ($n_{\mathrm{lost}}^{\mathrm{out}}$) are the photons lost at the input (output). We remark that the different distributions which yield to the same outcome in the lossy case present a significant total variation distance with respect to the lossless one (see Appendix C). Besides, the time required to classically simulate a lossy Scattershot BS event is a weighted average between the computation of the ${{n+n_{\mathrm{lost}}^{\mathrm{in}}}\choose {n_{\mathrm{lost}}^{\mathrm{in}}}}$ $n$ photons distributions when losses happen at the input and the ${{m-n+n_{\mathrm{lost}}^{\mathrm{out}}}\choose {n_{\mathrm{lost}}^{\mathrm{out}}}}$ possible evolutions for a $n-n_{\mathrm{lost}}^{\mathrm{out}}$ output. Note however that to simulate a $n-n_{\mathrm{lost}}^{\mathrm{out}}$ event we still need to evolve a $n$ photons state through the unitary.

\begin{figure}[ht!]
\centering
\includegraphics[width=0.6\textwidth]{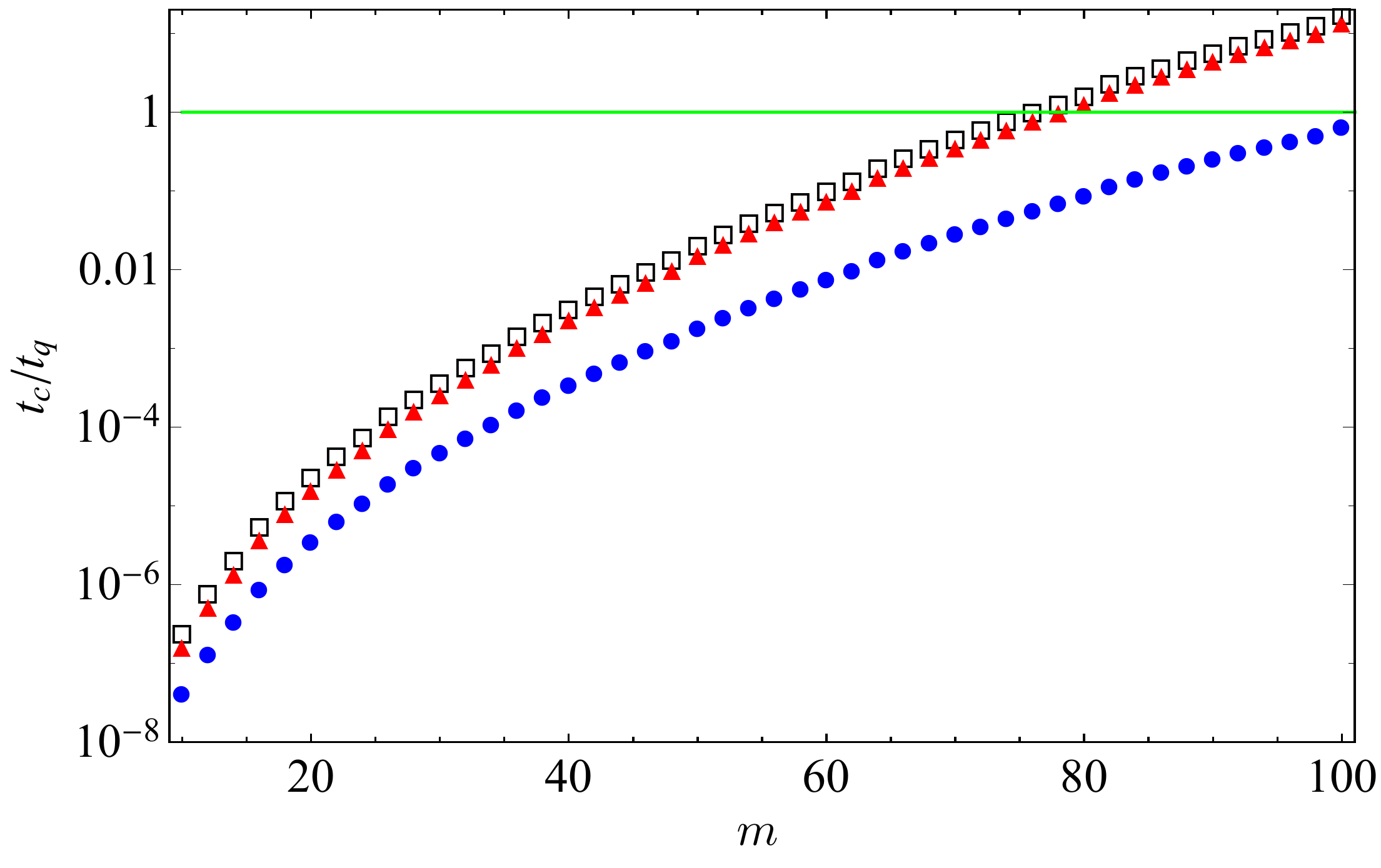}
\caption{Ratio between the time required to compute a single Scattershot BS event and to perform a single experimental run. Blue circles correspond to correct BS, red triangles identify the one lost photon case (equally likely at input and output) and black squares are the generalized BS (i.e. the sum of both). Experimental parameters are set as: $g=0.02$, $\eta_{\mathrm{T}}=0.6$, $p_{\mathrm{in}}=0.7$ and $\eta_{\mathrm{D}}=0.6-0.25*(m-10)/90$ (the chance that a photon crosses the whole interferometric network scales anti-linearly with the dimension). The plot is the result of the weighted average over all Scattershot BS events with $3\leq n < \sqrt{m}$.}
\label{qs_optical}
\end{figure}

We display in Fig. \ref{qs_optical} the results of the comparison between a classical and a quantum agent for a traditional Scattershot BS together with data of a case with constant losses. We vary the number of photons and sources and we look for all the $n$ photons events in accordance with the principle $n^2<m$. The detection efficiency is supposed to decrease when we increase the dimension of the optical network, since it includes the transition through the interferometer. Indeed, let us call $(1-p^{\mathrm{dc}}_l)$ the probability to lose a photon in an integrated beam splitter (a directional coupler) with current technology. The overall single-photon transmittivity then scales as $(p^{\mathrm{dc}}_l)^m$ for interferometer architectures where the number of beam-splitter layers scales as $m$. Assuming a feasible improvement in the experimental techniques to come alongside with the realization of larger devices, we obtain that the bound for quantum supremacy lies in a regime with $n_{\mathrm{th}} \lesssim 8$ photons and $m_{\mathrm{th}}\simeq 80$ sources and modes. Despite being experimentally demanding, this generalized scattershot BS reveals to be a step forward if compared with the previously estimated regime of $20$ photons in $400$ modes. In fact, on the one hand it requires a smaller interferometric network, less sensitive to losses, and on the other hand the lower number of photons increases the rate and loosens the requirements on the single photons and the optical elements fidelities \cite{Shchesnovich2015, Tillmann14, Leverrier2015}.

\section{Boson Sampling with quantum dot sources}

As we have highlighted in Secs. \ref{sec:SBS}-\ref{quantumsupremacybound}, the main issue of Boson Sampling with optical photons is the low scalability of SPDC sources due to the occurrence of multiple pairs events. Recent experiments have tried to overcome this problem relying on quantum dot sources \cite{Soma16,Lore16,He16}, where a train of single-photon pulses is deterministically generated (with up to $99\%$ fidelity) by a InGaAs quantum dot embedded in a micro-cavity and excited by a quasi-resonant laser beam \cite{Qdot1,Qdot2}. The emitted pulses are subsequently collected in a single-mode fiber with a total source efficiency $\eta$, which depends on the laser pump power due to saturation effects in the quantum dot. The most common approaches to convert a train of single photons equally separated in time in a $n$ single photon Fock state are passive \cite{Lore16} and active \cite{He16} demultiplexing. The former can be achieved by arranging a single array of $n-1$ beam splitters whose reflectivities and transmittivities are tuned such that the probability for each photon to escape the cascade is always $1/n$ (i.e. numbering the beam splitters from $1$ to $n-1$ their reflectivities scale as $1/(n-i+1)$). Maintaining the previous notation, the probability to successfully perform a BS experiments with $i\leq n$ photons injected from the first $i$ ports of the array reads

\begin{equation}\label{qdBS}
P^{\mathrm{BS}}_{\mathrm{QD}}(i)=\eta^i\frac{1}{n^i}p_{\mathrm{in}}^i\eta_\mathrm{D}^i.
\end{equation}

\begin{figure}[h!]
\centering
\includegraphics[width=0.99\textwidth]{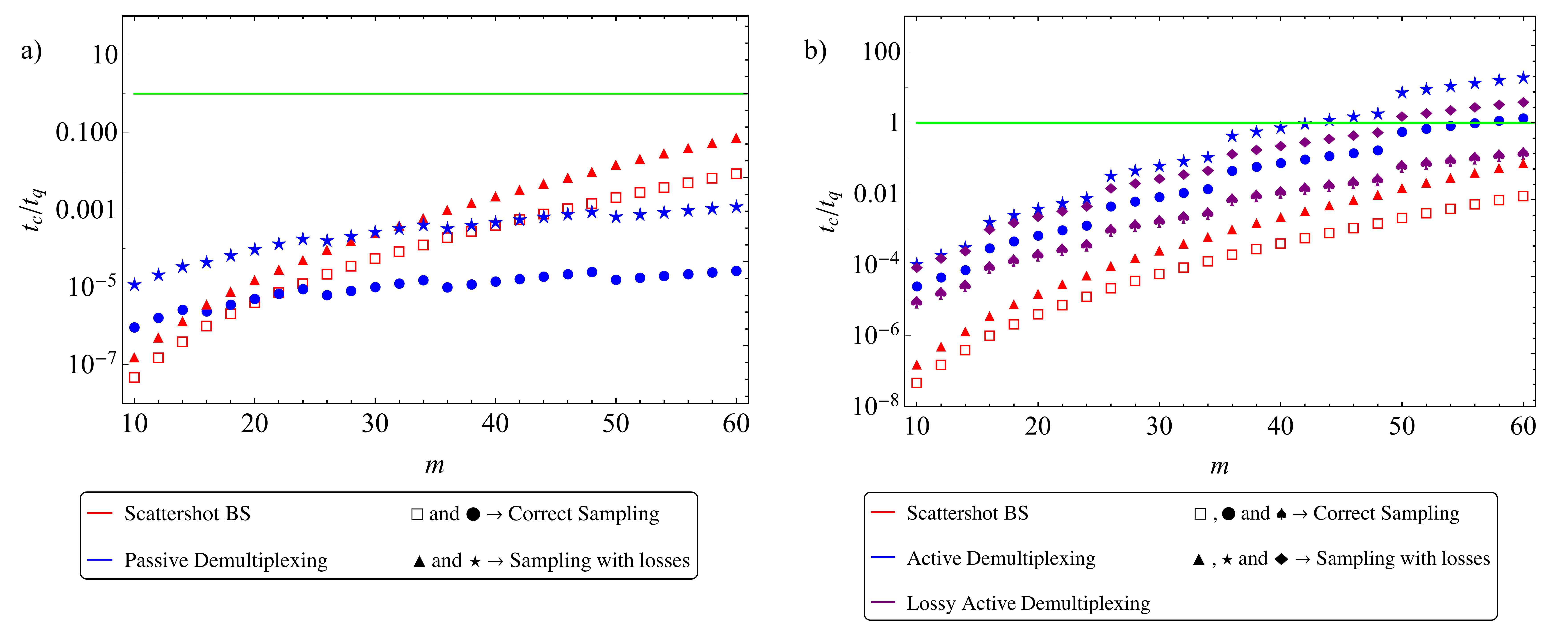}
\caption{Performances of the quantum dot source case by considering (a) a passive demultiplexing approach and (b) an active one, in comparison with heralded Scattershot Boson Sampling with SPDC sources (red points). (a) Blue points: passive demultiplexing. (b) Blue points: lossless active demultiplexing. Purple points: lossy active demultiplexing with efficiency $\eta_{\mathrm{dm}}=0.7$. Square, circle and spades depict a correct BS experiment, while triangle, star and rhomb points stand for the lossy case where a single photon is lost either at the input or at the output. Experimental parameters are set as: $\eta=0.35$, $p_{\mathrm{in}}=0.7$, $\eta_{\mathrm{T}}=0.6$, $p_{\mathrm{in}}=0.7$ and $\eta_{\mathrm{D}}=0.6-0.25*(m-10)/90$. The number of photons for the quantum dot case is increased in steps so as to fulfil the complexity requirement $m > n^2$, thus giving rise to the jumps appearing in the plot.}
\label{qdotsBS}
\end{figure}

We observe from Eq. (\ref{qdBS}) that Boson Sampling with quantum dots suffers a significant drawback when passive demultiplexing is adopted, since the probability of a successful event scales inversely with the factorial of the number of photons.  We plot in Fig. \ref{qdotsBS}a a comparison of the performances between Scattershot BS with SPDC single photon sources and BS with a quantum dot source and passive demultiplexing.
While the advantage is quite remarkable for a small number of photons, the adoption of passive demultiplexing reduces the efficiency for increasing $n$. A substantial improvement, proportional to $n^i$, can be achieved by exploiting an efficient active demultiplexing method, thus rendering quantum dot sources a promising platform to reach the quantum supremacy regime (see Fig. \ref{qdotsBS}b). In the latter case the probability of a successful BS run is supposed to scale as $P^{\mathrm{BS}}_{\mathrm{QD}}=\eta^i\eta_{\mathrm{dm}}^ip_{\mathrm{in}}^i\eta_\mathrm{D}^i$, where $\eta_{\mathrm{dm}}$ is the efficiency of the demultiplexing procedure (see Ref.\cite{xiong2016} for a technique with heralded photons)
\section{Boson Sampling with microwave photons}

Now that we have addressed the strengths and weaknesses of Scattershot BS for SPDC and quantum dot sources with optical photons, we discuss the expectations offered by a completely new approach \cite{huh2015bis}. Boson Sampling with microwave photons is a new experimental proposal that meets all the requirements (e.g. Fock states with indistinguishable photons in input, Haar random unitary transformation and entangled Fock states at the output) while carrying them out in a different way.
\begin{figure}[b!]
\centering
\includegraphics[width=0.6\textwidth]{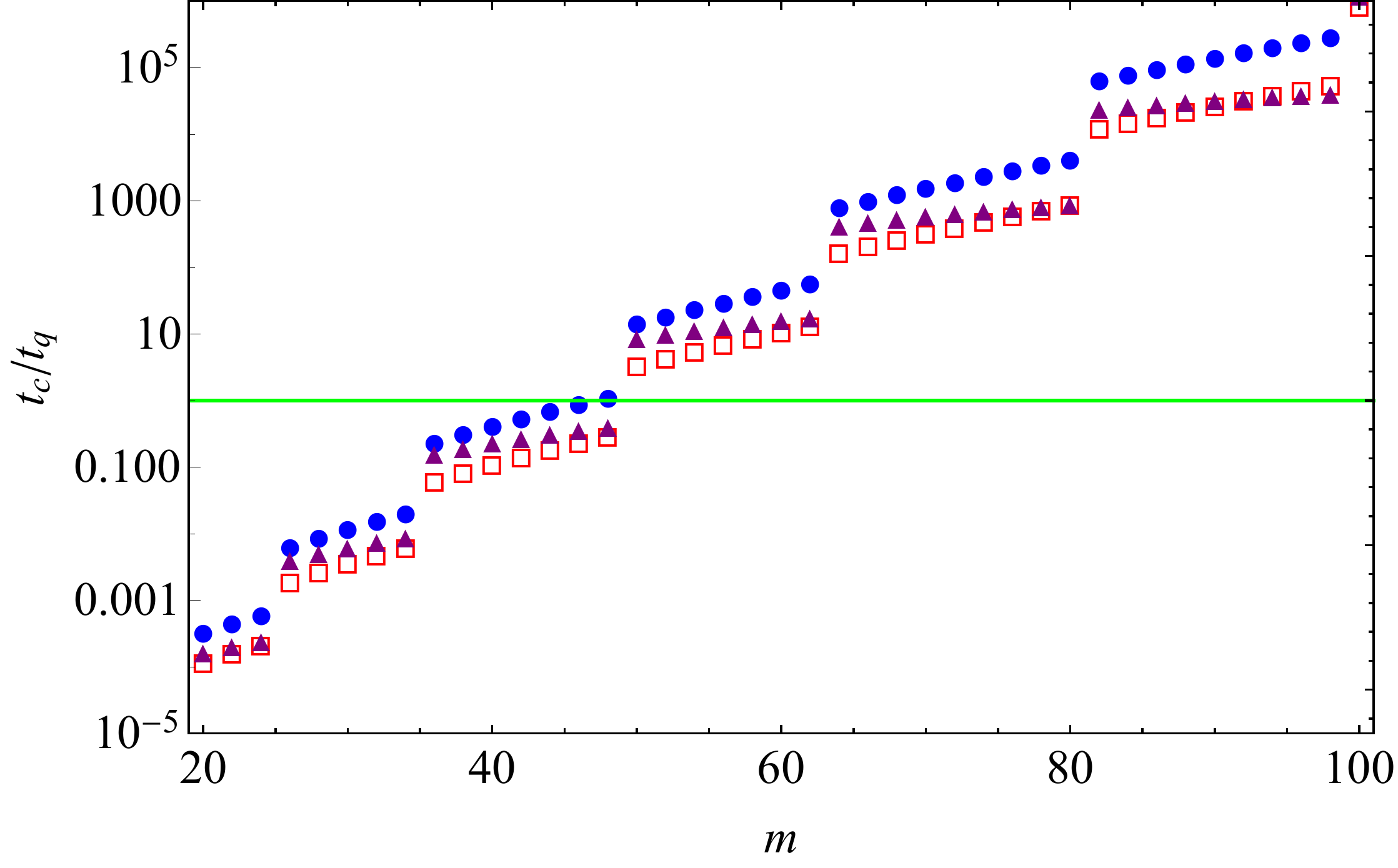}
\caption{Expected quantum supremacy bound for Boson Sampling with microwave photons. Red squares represent the case of a correct BS experiment, purple triangles consider the ratio $t_{\mathrm{c}}/t_{\mathrm{q}}$ when one photon is lost either at the input or at the output and there is a certain non-zero dark count probability ($p_{\mathrm{dark}}$), i.e. when photons are erroneously detected in vacuum modes. Blue dots represent the sum of the two cases. Experimental parameters are set as: $p_{\mathrm{dark}}=0.1$, $\eta_{\mathrm{D}}=0.7$, $p_{\mathrm{in}}=0.9$ and $t_{\mathrm{step}}= 0.3\mu s$. The number of photons is increased in steps so as to fulfil the complexity requirement $m> n^2$, thus giving rise to the jumps appearing in the plot.}
\label{microBS}
\end{figure}
More specifically, $n$ photons are deterministically generated by exciting $n$ X-mon qubits among a chain composed by $m$ qubits (potentially at very high repetition rate $\sim 10^2 \mathrm{MHz}$), each coupled both to a storage and a measurement resonator through a Jaynes-Cummings interaction. By tuning their frequencies through an external magnetic field the $n$ selected qubits are set in resonance with the storage resonator, thus creating $n$ single photon Fock states with high efficiency \cite{hofheinz2008, hofheinz2009}. The interferometric network of beam splitters and phase shifters implementing the $m\times m$ unitary transformation is replaced by the chain of time-dependently interacting cavities. A superconducting ring with a Josephson junction is used to tune the coupling between resonators $i$ and $i+1$ and acts as a beam splitter \cite{peropadre2013} described by the interaction Hamiltonian $H_{\mathrm{int}}=g_{i,i+1}(a_i a_{i+1}^\dag +h.c.)$, with $g_{i,i+1}\sim 50 \mathrm{MHz}$. Lasting only $t_{\mathrm{bs}}\sim 0.02 \mu s$, this interaction can be turned on and off very rapidly (in the order of nanoseconds) and has already reported high coupling ratios $O(10^4)$ \cite{baust2015, wulschner2015}.
Meanwhile, a phase shift operation can be realized by exciting the qubit associated with the cavity and pushing it off resonance for a time $t_{\mathrm{ps}}$, inducing a frequency shift among the resonators that after an appropriate time is turned into a phase shift. By subsequently applying beam splitter and phase shifter operations a $m\times m$ unitary transformation can be implemented after $O(m)$ steps \cite{Reck94}, each requiring $t_{\mathrm{step}}=t_{\mathrm{ps}}+t_{\mathrm{bs}}\sim 0.3\mu s$. Considering that the typical cavity decoherence time is around $\tau=1/\kappa\sim 100 \mu s$ \cite{wang2013, ohya2014}, this potentially permits to perform more than one hundred steps (assuming a sufficiently high fidelity for each operation). Eventually, the input preparation procedure is inverted to measure the output state: the qubits are put in resonance with the cavities and through a Jaynes-Cummings interaction those whose corresponding resonators contain a photon are naturally excited \cite{hofheinz2008}. A non demolition measurement of the qubit state can finally be addressed with more than 90\% efficiency by coupling it with a low quality cavity \cite{johnson2010}. Even though in the regime $m=O(n^2)$ (where BS was proved to be hard) attention can be restricted to those outcomes with 0 or 1 photon for each output due to the \textit{Boson Birthday Paradox} \cite{Aaronson10,Spagnolo2013}. Furthermore, superconducting qubits are a promising tool towards generalizations employing photon number resolving detectors \cite{schuster2007}.

Using feasible experimental parameters provided in Ref. \cite{huh2015bis}, we can evaluate the threshold $t_{\mathrm{c}}/t_{\mathrm{q}}>1$ for a generalized version of BS with microwave photons. The calculations are similar to the ones for the case of optical photons without the issue of double pairs generation, but with the only constraint that the experimental rate is bounded by $ (m\times 0.3 \, \mu s)^{-1}$, scaling inversely with the number of steps (modes) $m$. We refer to Appendix D for the complete expressions that take into account also losses and dark counts (which have been proved to be theoretically equivalent to losses at the input \cite{Aaronson2015}). We present a summary of the results in Fig. \ref{microBS}, where the ratio $t_{\mathrm{c}}/t_{\mathrm{q}}$ is plotted as a function of the number of modes $m$, i.e. the dimension of the unitary. In this case we keep the detection efficiency $\eta_{\mathrm{D}}$ constant, since we expect negligible losses of photons for times quite below the cavities decoherence time. The final theoretical result leads to a significant improvement in the efficiency and an additional step towards quantum supremacy which can be achieved with a $7$ photons in $50$ modes experiment.

\section{Conclusions}

We have reviewed the problem of Boson Sampling together with its most recent extensions and variations: Scattershot and lossy sampling and the proposal to adopt photons in the microwave spectrum. In particular, we have highlighted the strengths and weaknesses of the model under reasonable experimental assumptions in order to understand if and how BS can be an effective approach to assess quantum supremacy. Using SPDC sources for single photons generation has the unavoidable drawback of multiple pairs, hence leading to experimentally inaccurate results that worsen by increasing the number of photons. Besides, not only the generation of an initial state with large $n$ can be hardly achieved, but it also requires higher mutual fidelities among the particles and higher accuracy for the optical elements to preserve the scalability. Recent experimental results on quantum dot sources \cite{Soma16,Lore16,He16} can open the way to new perspectives in the implementation of Boson Sampling with large photon numbers, due to their high generation efficiency and high photon indistinguishability.

Performing a state-of-the-art analysis, we have shown that the threshold $t_{\mathrm{c}}>t_{\mathrm{q}}$, i.e. the regime where the quantum agent samples faster (in time $t_{\mathrm{q}}$) than his classical counterpart (in time $t_{\mathrm{c}}$), can be achieved with a Scattershot BS experiment with $m_{\mathrm{th}}\simeq 80$ SPDC sources, far less than the original regime of $n_{\mathrm{th}}=20$ photons and $m_{\mathrm{th}}=400$ modes. While on the one hand the permanent guarantees the complexity of the problem, on the other hand it eventually reveals to be much more convenient to increase the sampling rate, rather than focusing on the size of the permanent. Indeed, a crucial role to reach the $t_{\mathrm{c}}>t_{\mathrm{q}}$ regime is played by the disposal of many sources in parallel, together with the possibility of sampling every time from a different random input and, most of all, the inclusion of events with constant losses of photons. The same analysis conducted with quantum dot sources and an active demultiplexing approach reported a theoretical attainment of the bound with $n_{\mathrm{th}}\simeq 7$ photons and $m_{\mathrm{th}}\simeq 50$ modes.

Aiming to maximize the efficiency and the accuracy of the protocol, we have finally analyzed a new suggestion by Peropadre et al. \cite{huh2015bis} consisting in the adoption of on demand microwave photons. This proposal overcomes the problem of erroneous input states and provides a remarkable decrease of losses, thus enhancing the experimental rate. Since in this case the unitary is implemented in time (it is decomposed in a number of steps performed one after the other), the sampling rate scales inversely with the dimension. However, this does not constitute a relevant issue considering that the bound for quantum supremacy is lowered to $n_{\mathrm{th}}\simeq 7$ photons in $m_{\mathrm{th}}\simeq 50$ modes, which requires a running time appreciably below the decoherence time of the system.

\section*{Acknowledgements}
The authors acknowledge E. F. Galvao, D. J. Brod and J. Huh for very useful discussions on the subject of this paper. This work was supported by the H2020-FETPROACT-2014 Grant QUCHIP (Quantum Simulation on a Photonic Chip; grant agreement no. 641039, http://www.quchip.eu).

\appendix

\section{SPDC sources}
The two-mode SPDC state reads $\vert \Psi^{\mathrm{SPDC}} \rangle = \sum_s \lambda_s\ \vert s,s \rangle$, the related photon number probability distribution being
\begin{equation}
\label{spdc}
P^{\mathrm{SPDC}}(s)=\left|\lambda_s\right|^2=\frac{\tanh^{2s}\chi}{\cosh^2\chi},
\end{equation}
where $s$ is the number of photons per mode and $\chi$ is the squeezing parameter.
Since quantum supremacy is expected to require quite a large number of possible input states and sources, it is mandatory to evaluate the contribution of second order generation terms. If we define $g=P^{\mathrm{SPDC}}(1)$ as the probability of generating a single pair, then from Eq. (\ref{spdc}) the second order scales as $P^{\mathrm{SPDC}}(2)\sim g^2$.

\section{Boson Sampling from erroneous input state}
Given an apparently correct sample with $n$ photons triggered at the input and output, this could be the result of the injection of couples of photons such that even though the total number of particles is $n$, the effective input state is different from the expected one. The probability of injecting $n_1$ single photons while triggering $n$ entries results to be
\begin{eqnarray}
\fl 
P_{\mathrm{in}}^{(\mathrm{fake})}(n,n_1)=\sum_{2x+y+z\ge n \atop x+y+z\le n}p_{\mathrm{in}}^{z}(1-p_{in})^{n_1-z}{n_1 \choose z} p_{\mathrm{in}}^{2x}[2p_{\mathrm{in}}(1-p_{\mathrm{in}})]^{y}(1-p_{\mathrm{in}})^{2(n-n_1-x-y)} \nonumber \\ 
\times {{n-n_1} \choose {x,y}}=
\sum_{x=1}^{n-n_1}\sum_{w=n-2x}^{n-x}\sum_{z=\frac{n-2x-w}{2}}^{n_1}2^{w-z}p_{\mathrm{in}}^{w+2x}(1-p_{\mathrm{in}})^{2n-n_1-2x-w} \nonumber\\ \times {n_1 \choose z}{{n-n_1} \choose {x,w-z}},
\end{eqnarray}
where $w=z+y$ is the total number of injected single photons, $z$ and $y$ being the fractions coming respectively from single photons and pairs impinging on the input ports of the interferometer. The outer sum has the constraints $x+y+z\le n$ (otherwise we would trigger $n'\ge n$ photons in input) and $2x+y+z\ge n$ (otherwise we would detect less than $n$ photons at the end of the chip), being $x$ the number of erroneously injected pairs.

In particular, if we generate $s$ single photons and $t$ pairs, the probability to inject a fake state ($0\leq n_1 < n$ single photons and $n-n_1$ pairs), despite having heralded an apparently correct one, and detect $n$ photons at the output is given by
\begin{eqnarray}
\fl P_{\mathrm{trig,det}}^{(\mathrm{fake})}(n|s,t)=P_{\mathrm{gen}}^{(2)}(s,t)\sum_{n_1=\max[n-t,0]}^{\min[s,n]}P_{\mathrm{in}}^{(\mathrm{fake})}(n,n_1)\eta_{\mathrm{T}}^{n_1}\eta_{\mathrm{T}_{2}}^{n-n_1}(1-\eta_{\mathrm{T}})^{s-n_1} {s \choose n_1} \nonumber\\
\times (1-\eta_{\mathrm{T}_{2}})^{t-n+n_1}{{t} \choose {n-n_1}} \eta_{D}^{n} (1-\eta_{D})^{n-n_{1}} {2n-n_{1} \choose n}.
\end{eqnarray}
Here the sum considers all possible cases with at least an extra injected photon coming from a double pair ($n-n_{1}$), where $n$ trigger detectors successfully click ($n_{1}$ with single-photon input and $n-n_{1}$ with two-photon inputs, detection efficiencies $\eta_{\mathrm{T}}$ and $\eta_{\mathrm{T}_{2}}$ respectively), and $n$ photons are detected at the output (efficiency $\eta_{D}$). We remark that we did not distinguish those events in which a greater number of photons is injected in the interferometer, some of them get lost within the chip and finally only $n$ are successfully detected. Indeed, we included all kind of losses (i.e. those in the interferometer and those at the detection stage) in the parameter $\eta_{\mathrm{D}}$. The effect of losses during the unitary evolution on the Boson Sampling distribution would be very subtle to estimate, and might be addressed in future works; however, it does not affect our aim of finding a \emph{lower} bound for quantum supremacy.

\section{Validation Data}
We report hereafter additional simulated data on the validation for $m=40$ modes lossy Boson Sampling against the distinguishable sampler in the case of losses at the input, at the output and both (see tables C.1, C.2 and C.3 respectively). We finally sum up the most relevant cases.
The probability assigned to each event of a lossy BS distribution is obtained by averaging over all possible samplings that could have led to the given lossy outcome. More specifically, when $n_{\mathrm{lost}}^{\mathrm{in}}$ photons are lost at the input and $n$ are detected at the output we will have to mediate over ${{n+n_{\mathrm{lost}}^{\mathrm{in}}}\choose {n_{\mathrm{lost}}^{\mathrm{in}}}}$ distributions, each corresponding to a possible input state with $n$ photons. The same applies when we know that losses occur before the output detection: this time we will have ${{m-n+n_{\mathrm{lost}}^{\mathrm{out}}}\choose {n_{\mathrm{lost}}^{\mathrm{out}}}}$ possible distributions to weight. Conversely, if we assume that photons can be lost at the input and at the output we will have to average over all events such that $n_{\mathrm{lost}}^{\mathrm{in}}+n_{\mathrm{lost}}^{\mathrm{out}}=n_{\mathrm{lost}}$. This means that for every value of $n_{\mathrm{lost}}^{\mathrm{in}}$ we mediate over the combination of possible inputs to reconstruct the theoretical output distribution from which in turn we deduce the probability for $n_{\mathrm{her}}-n_{\mathrm{lost}}$ photons events (where $n_{\mathrm{her}}=n+n_{\mathrm{lost}}^{\mathrm{in}}$ is the number of heralded input photons). 

\begin{table}[ht!]
\centering
{\small
\begin{tabular}{cccc}
Photons [$n$] & Losses [$n_{\mathrm{lost}}^{\mathrm{in}}$] & Inputs & \# samples\\
\hline
3 & 0 & 1  & $19\pm 3$\\
3 & 1 & 4  & $50\pm 6$\\
3 & 2 & 10 & $88\pm 9$ \\ 
4 & 0 & 1  & $14\pm 2$\\
4 & 1 & 5  & $37\pm 4$\\
4 & 2 & 15 & $64\pm 6$\\
5 & 0 & 1  & $12\pm 2$\\
5 & 1 & 6  & $31\pm 3$\\
5 & 2 & 21 & $54\pm 6$\\
6 & 0 & 1  & $11\pm 1$\\
6 & 1 & 7  & $29\pm 3$\\
\end{tabular}
}
\caption{Minimum data set size to validate BS with losses occurring at the input against a sampling with distinguishable photons with a 95\% confidence level. The results have been averaged over 100 Haar random $40\times 40$ unitaries. Inputs indicate the number of possible inputs combinations, given the number of injected photons and losses.}
\label{var_err_dist_1}
\end{table}

\begin{table}[ht!]
\centering
{\small
\begin{tabular}{ccccc}
Photons [$n$] & Losses [$n_{\mathrm{lost}}^{\mathrm{out}}$] & Outputs & \# samples\\
\hline
3 & 0 & 1  & $19\pm 2$\\
3 & 1 & 38  & $101\pm 14$\\
4 & 0 & 1  & $14\pm 2$\\
4 & 1 & 37  & $53\pm 6$\\
4 & 2 & 703 & $208\pm 24$\\
5 & 0 & 1  & $12\pm 2$\\
5 & 1 & 36  & $41\pm 4$\\
5 & 2 & 66 & $109\pm 12$\\
\end{tabular}
}
\caption{Minimum data set size to validate BS with losses occurring at the output against a sampling with distinguishable photons with a 95\% confidence level. The results have been averaged over 100 Haar random $40\times 40$ unitaries. Outputs indicate the number of possible $n$ photons output combinations from which a generic sampled event could come from.}
\label{var_err_dist_2}
\end{table}

\begin{table}[ht!]
\centering
{\small 
\begin{tabular}{cccccccc}
Modes [$m$] & Photons [$n$] & Losses [$n_{\mathrm{lost}}$] & \# samples (in) & \# samples (out) & \# samples (both)\\
\hline
30 & 3 & 1 & $95\pm 12$  & $103\pm 13$ & $179\pm 35$\\
 & 4 & 1 & $52\pm 6$  & $56\pm 7$ & $80\pm 10$\\
 & 5 & 2 & $95\pm 11$  & $121\pm 13$ & -\\
 & 5 & 1 & $38\pm 4$  & $43\pm 6$ & $55\pm 6$\\
 & 6 & 2 & $69\pm 7$  & $93\pm 9$ & -\\
 & 6 & 1 & $33\pm 4$  & $39\pm 4$ & $49\pm 5$\\
 & 7 & 2 & $59\pm 7$  & $86\pm 7$ & -\\
40 & 3 & 1 & $93\pm 12$  & $101\pm 14$ & $181\pm 31$\\
 & 4 & 1 & $50\pm 6$  & $53\pm 6$ & $78\pm 8$\\
 & 5 & 2 & $88\pm 9$  & $109\pm 12$ & -\\
 & 5 & 1 & $37\pm 4$  & $41\pm 4$ & $52\pm 7$\\
\end{tabular}
}
\caption{Minimum data set size to validate BS with losses occurring either at the input, at the output or both cases against a sampling with distinguishable photons with a 95\% confidence level. The results have been averaged over 100 Haar random $40\times 40$ unitaries. The number of photons indicates how many photons are actually detected.}
\label{var_err_dist_3}
\end{table}

\newpage

We finally report in table C.4 the simulated error distance of Boson Sampling distributions concerning different possible input states ($\nu_{\mathrm{err}}=1/2\sum_i|p(i)-p^\prime(i)|$, where $p(i)$ and $p^\prime(i)$ are the probabilities assigned to event $i$ by the two distributions and the sum is intended over all possible events). In particular, we compare the single $n$ photon Fock state distributions to some other inputs, respectively with second order terms, vacuum states and an extra photon that is lost at the detection. 

For example, considering the case $n=3$, we compare the correct input 1-1-1 (all other $m-n$ entries are $0$) with input states 2-1-0 (the fourth photon is triggered but not injected), 2-1-1 (there is an extra pair but one photon is not detected), 1-1-0-1 (1-1-1-1 is generated and one photon is lost at the input) and 1-1-1-1 (one photon is lost at the detection). We report an overview of the variational error distance for several values of the number of photons $n$ and modes $m$.
\begin{table}[ht!]
\centering
{\small
\begin{tabular}{cccccc}
Photons [$n$] & Modes [$m$] & 2-1-0 & 2-1-1 & 1-0-1-1 & 1-1-1-1\\
\hline
3 & 15 & $0.473\pm 0.054$ & $0.222\pm 0.016$ & $0.316\pm 0.026$ & $0.337\pm 0.029$\\
  & 25 & $0.460\pm 0.036$ & $0.210\pm 0.013$ & $0.318\pm 0.018$ & $0.326\pm 0.020$\\
  & 40 & $0.466\pm 0.031$ & $0.211\pm 0.011$ & $0.321\pm 0.018$ & $0.327\pm 0.015$\\
  & 50 & $0.444\pm 0.024$ & $0.205\pm 0.013$ & $0.320\pm 0.015$ & $0.315\pm 0.013$\\ 
4 & 15 & $0.482\pm 0.041$ & $0.263\pm 0.013$ & $0.328\pm 0.022$ & $0.349\pm 0.018$\\
  & 20 & $0.467\pm 0.032$ & $0.251\pm 0.011$ & $0.333\pm 0.019$ & $0.346\pm 0.019$\\
  & 30 & $0.464\pm 0.021$ & $0.245\pm 0.009$ & $0.329\pm 0.014$ & $0.332\pm 0.010$\\
5 & 20 & $0.473\pm 0.023$ & $0.276\pm 0.010$ & $0.338\pm 0.015$ & $0.353\pm 0.011$\\
  & 25 & $0.475\pm 0.027$ & $0.267\pm 0.008$ & $0.336\pm 0.012$ & $0.350\pm 0.011$\\
  & 30 & $0.461\pm 0.010$ & $0.256\pm 0.008$ & $0.336\pm 0.012$ & - \\
6 & 25 & $0.468\pm 0.018$ & $0.280\pm 0.007$ & $0.341\pm 0.012$ & - \\
\end{tabular}
}
\caption{Variational error distance averaged over 100 samples between the ideal Boson Sampling distribution with $n$ single input photons in the state 1-$\cdots$-1 and some wrong samplings. The state 2-1-0 (2-1-1) generically represents the 2-1-$\cdots$-1-0 (2-1-$\cdots$-1) state with a couple of photons and $n-2$ ($n-1$) remaining single photons (e.g. 2-1-1-0 and 2-1-1-1 for $n=4$). The error distance with respect to the state 1-1-1-1 tells how far BS with $n$ photons is from BS with $n+1$ photons, one of which is lost at the output.}
\label{var_err_dist_4}
\end{table}
To better understand the significance of the values in the table we computed the variational error distance for distributions with a completely wrong input state. Always with respect to the 1-1-1 case, we got for $n=3$ photons in $m=50$ modes: 3-0-0 $\rightarrow (0.593 \pm 0.050)$, 0-0-0-3 $\rightarrow 0.699 \pm 0.050)$ and 0-0-0-1-1-1 $\rightarrow (0.630 \pm 0.035)$.

\section{Boson Sampling with microwave photons}
If we call $p_{\mathrm{in}}$ the probability to successfully excite a X-mon qubit and then create a single photon in the coupled cavity through a Jaynes-Cummings interaction, then the probability to lose $n_i$ photons at the input is
\begin{equation}
P_{\mathrm{MW}}^{\mathrm{in}}(n,n_{\mathrm{lost}}^{\mathrm{in}})=p_{\mathrm{in}}^{n-n_{\mathrm{lost}}^{\mathrm{in}}}(1-p_{\mathrm{in}})^{n_{\mathrm{lost}}^{\mathrm{in}}}{{n} \choose {n_{\mathrm{lost}}^{\mathrm{in}}}}.
\end{equation}
From this quantity we can evaluate the probability to perform a microwave BS losing $n_{\mathrm{lost}}$ photons overall (i.e these losses can occur either at the input or at the output)
\begin{equation}
\fl P_{\mathrm{MW}}^{\mathrm{lossy}}(n,n_{\mathrm{lost}})=\sum_{n_{\mathrm{lost}}^{\mathrm{in}}=0}^{n_{\mathrm{lost}}}[\eta_{\mathrm{D}}^{n-n_{\mathrm{lost}}}(1-\eta_{\mathrm{D}})^{n_{\mathrm{lost}}-n_{\mathrm{lost}}^{\mathrm{in}}}{{n-n_{\mathrm{lost}}^{\mathrm{in}}}\choose {n_{\mathrm{lost}}-n_{\mathrm{lost}}^{\mathrm{in}}}}P_{\mathrm{MW}}^{\mathrm{in}}(n,n_{\mathrm{lost}}^{\mathrm{in}}).
\end{equation}
This result assumes that dark counts are negligible (the chance to erroneously detect photons ($p_{\mathrm{d}}$) in vacuum modes is very low). If this is not the case, the probability for a microwave BS with $n_{\mathrm{lost}}$ photons becomes
\begin{eqnarray}
\fl P_{\mathrm{MW}}^{\mathrm{lossy}+\mathrm{dark}}(m,n,n_{\mathrm{lost}})=\sum_{n_{\mathrm{lost}}^{\mathrm{in}}=0}^{n_{\mathrm{lost}}}\sum_{j=0}^{n-n_{\mathrm{lost}}} \eta_{\mathrm{D}}^{n-n_{\mathrm{lost}}-j} (1-\eta_{\mathrm{D}})^{n_{\mathrm{lost}}+j-n_{\mathrm{lost}}^{\mathrm{in}}}{{n-n_{\mathrm{lost}}^{\mathrm{in}}}\choose{n_{\mathrm{lost}}-n_{\mathrm{lost}}^{\mathrm{in}}}} \nonumber\\
\times {{n-n_{\mathrm{lost}}}\choose{j}} p_{\mathrm{d}}^j(1-p_{\mathrm{d}})^{m-n+n_{\mathrm{lost}}^{\mathrm{in}}}{{m-n+n_{\mathrm{lost}}^{\mathrm{in}}+j}\choose{j}} P_{\mathrm{MW}}^{\mathrm{in}}(n,n_{\mathrm{lost}}^{\mathrm{in}}).
\end{eqnarray}

\section*{References}

\bibliographystyle{iopart-num}
\bibliography{BosonSampling}

\end{document}